\documentclass[11pt,a4paper]{article}
\usepackage{amsmath,amssymb}
\usepackage{epsfig,graphicx}
\graphicspath{{ps/}}

\begin{document}

\title{\bf Recent results from the Belle experiment}

\author{
T.A.-Kh.~Aushev$^{\,a,b}$\footnote{{\bf e-mail}: aushev@itep.ru}\\
for the Belle Collaboration\\
$^a$\small{\em Swiss Federal Institute of Technology of Lausanne, EPFL}\\
$^b$\small{\em Institute for Theoretical and Experimental Physics, Moscow, ITEP}\\
}
\date{}
\maketitle

\begin{abstract}
We report the recent results of a search for the decay
$B^-\to\tau^-\bar\nu_\tau$, observations of new resonances $X,Y$ and
$Z$, and the first results from $\Upsilon(5S)$ data collected with the
Belle detector at KEKB $e^+e^-$ collider.
\end{abstract}

\section{Belle detector and KEKB collider}

The results reported in this paper were obtained using the data
collected with the Belle detector~\cite{beldetec} at the KEKB
asymmetric-energy $e^+e^-$ ($3.5$ on $8.0${\rm~GeV})
collider~\cite{KEKB} operating at the $\Upsilon(4S)$ resonance ($\sqrt
s=10.58\,{\rm GeV}$).  The Belle detector is a large-solid-angle
magnetic spectrometer that consists of a silicon vertex detector
(SVD), a 50-layer central drift chamber (CDC), a mosaic of aerogel
threshold \v{C}erenkov counters (ACC), time-of-flight scintillation
counters (TOF), and an array of CsI(Tl) crystals (ECL) located inside
a superconducting solenoid coil that provides a $1.5$~T magnetic
field.  An iron flux-return located outside of the coil is
instrumented to detect $K_L$ mesons and to identify muons (KLM).

Charged tracks are reconstructed in the CDC and their impact
parameters are precisely determined using the SVD.  Charged hardon
identification is accomplished based on the combined information from
the ACC, TOF and CDC $dE/dx$ systems.  Electron identification is
based on a combination of $dE/dx$ measurements, the ACC response and
information about the shape, energy deposit and position of the
associated shower in the ECL. Muons are identified by requiring an
association between KLM hits and an extrapolated track. Photons are
reconstructed in the ECL as showers that are not associated with
charged tracks.

\section{\boldmath Search for $B^-\to\tau^-\bar\nu_\tau$}

\begin{figure}[htb]
\centering\includegraphics[width=.5\textwidth]{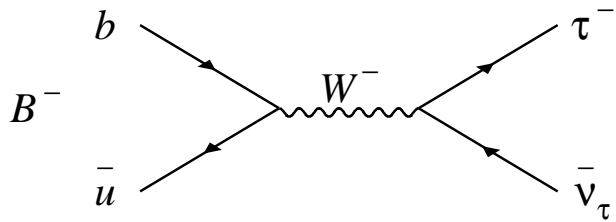}
\caption{Purely leptonic $B$ decay proceeds via quark annihilation
into a $W$ boson.}
\label{tau_diag}
\end{figure}

In the Standard Model (SM), the purely leptonic decay
$B^+\to\tau^+\nu_\tau$~\cite{cc} proceeds via annihilation of $b$ and
$\bar u$ quarks into a $W^-$ boson (Fig.~\ref{tau_diag}).
It provides a direct determination of the product of the $B$ meson
decay constant $f_B$ and the magnitude of the
Cabibbo-Kobayashi-Maskawa matrix element $|V_{ub}|$.  The branching
fraction is given by
\[
{\cal B}(B^-\to\tau^-\bar\nu_\tau)=\frac{G_F^2m_Bm_\tau^2}{8\pi}
(1-\frac{m_\tau^2}{m_B^2})^2f_B^2|V_{ub}|^2\tau_B,
\]
where $G_F$ is the Fermi coupling constant, $m_B$ and $m_\tau$ are the
$B$ and $\tau$ masses, and $\tau_B$ is the $B^-$ lifetime~\cite{PDG}.
The expected branching fraction is $(1.59\pm0.40)\times10^{-4}$,
obtained using $|V_{ub}|=(4.39\pm0.33)\times10^{-3}$, determined with
inclusive charmless semileptonic $B$ decay data~\cite{tau_ref1},
$\tau_B=1.643\pm0.010$~ps~\cite{tau_ref1}, and $f_B=0.216\pm0.022$~GeV
obtained from lattice QCD calculations~\cite{tau_ref2}.  Physics
beyond the SM, such as supersymmetry or two-Higgs doublet models,
could modify ${\cal B}(B^-\to\tau^-\bar\nu_\tau)$ through the
introduction of a charged Higgs boson~\cite{tau_ref3}.  Purely
leptonic $B$ decays have not been observed in past experiments.  The
most stringent upper limit on $B^-\to\tau^-\bar\nu_\tau$ comes from
the BABAR experiment: ${\cal
B}(B^-\to\tau^-\bar\nu_\tau)<2.6\times10^{-4}$ (90\%
C.L.)~\cite{tau_ref4}.

To reconstruct this decay mode, one $B$ meson was fully reconstructed
in the event (referred to hereafter as the tag side ($B_{\rm tag}$)),
and the properties of the remaining particle(s) (referred to as the
signal side ($B_{\rm sig}$)), are compared to those expected for
signal and background.  The method allows to suppress strongly the
combinatorial background from both $B\bar B$ and continuum events.

In the events where a $B_{\rm tag}$ is reconstructed, decays of
$B_{\rm sig}$ into a $\tau$ and a neutrino were searched for.
Candidate events are required to have one or three charged track(s) on
the signal side with total charge opposite to that of $B_{\rm tag}$.
The $\tau$ lepton is identified in the five decay modes
$\mu^-\bar\nu_\mu\nu_\tau$, $e^-\bar\nu_e\nu_\tau$, $\pi^-\nu_\tau$,
$\pi^-\pi^0\nu_\tau$ and $\pi^-\pi^+\pi^-\nu_\tau$, which taken
together correspond to $81\%$ of all $\tau$ decays.

\begin{figure}[htb]
\centering\includegraphics[width=.35\textwidth]{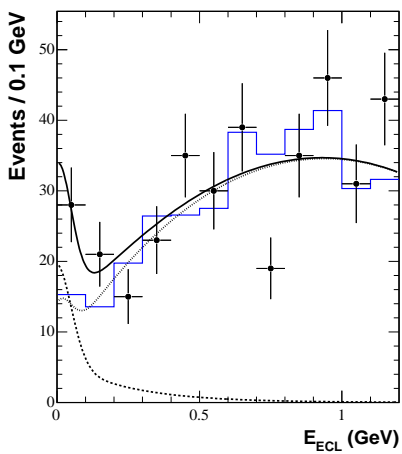}
\caption{$E_{\rm ECL}$ distributions in the data.  The data and
background Monte Carlo (MC) samples are represented by the points with
error bars and the solid histogram, respectively.  The solid curve
shows the result of the fit with the sum of signal shape (dashed) and
background shape (dotted).}
\label{tau_signal}
\end{figure}

The most powerful variable for separating signal and background is the
remaining energy in the ECL, denoted as $E_{\rm ECL}$ and defined as
the sum of the energy deposits in the ECL that are not associated with
either the $B_{\rm tag}$ or the $\pi^0$ candidate from the
$\tau^-\to\pi^-\pi^0\nu_\tau$ decay.  For signal events, $E_{\rm ECL}$
must be either zero or a small value arising from beam background
hits.  Therefore, signal events peak at low $E_{\rm ECL}$.  On the
other hand background events are distributed toward higher $E_{\rm
ECL}$ values due to the contribution of additional neutral clusters.
Figure~\ref{tau_signal} shows the obtained $E_{\rm ECL}$ distribution
when all $\tau$ decay modes are combined.  One can see a significant
excess of events in the $E_{\rm ECL}$ signal region below $E_{\rm
ECL}<0.25$~GeV.  The number of signal events in the signal region
deduced from the fit is $17.2^{+5.3}_{-4.7}$.  The obtained branching
fraction is $(1.79^{+0.56}_{-0.49}({\rm stat})\,^{+0.46}_{-0.51}({\rm
syst}))\times10^{-4}$.  The significance of the observed signal is
$3.5\,\sigma$.

Using the measured branching fraction and the known values of $G_F$,
$m_B$, $m_\tau$, $\tau_B$ and $|V_{ub}|$, one obtains
$f_B=0.229^{+0.036}_{-0.031}({\rm stat})^{+0.034}_{-0.037}({\rm
syst})$~GeV, which is the first determination of the $B$ meson decay
constant~\cite{tau_ref5}.

\section{Observation of new resonances}

Recently there has been a revival of interest in the possible
existence of mesons with a more complex structure than the simple
$q\bar q$ bound state of the original quark model.  There are
long-standing predictions of four-quark $q\bar qq\bar q$ meson-meson
resonance states~\cite{xyz_ref1} and for $q\bar q-gluon$ hybrid
states~\cite{xyz_ref2}.  Searches for such particles in systems
including a charmed-anticharmed quark pair ($c\bar c$) are
particularly effective because, for at least some of these cases, the
states are expected to have clean experimental signatures as well as
relatively narrow widths, thereby reducing the possibility of overlap
with standard $c\bar c$ mesons.

\subsection{\boldmath Observation of $X(3872)$}

The Belle experiment discovered a new state, named $X(3872)$, as a
narrow peak in the $\pi^+\pi^-J/\psi$ mass spectrum from exclusive
$B\to K\pi^+\pi^-J/\psi$ decays~\cite{x3872_ref1}.  This observation
has been confirmed by other experiments~\cite{x3872_ref2}.  The
properties of the $X(3872)$ do not match well to any $c\bar c$
charmonium state~\cite{x3872_ref3}.  This, together with the close
proximity of the $X(3872)$ mass with the $m_{D^0}m_{D^{*0}}$ mass
threshold, have led some authors to interpret the $X(3872)$ as a
$D^0\bar D^{*0}$ resonant state~\cite{x3872_ref4}.
                                                          
To investigate further and determine its quantum numbers, the
$X(3872)$ state was searched in various decay channels.  It was
observed in two decay modes: $X(3872)\to\gamma J/\psi$ from $B^+\to
K^+\gamma J/\psi$ and $X(3872)\to\pi^+\pi^-\pi^0J/\psi$ from $B^+\to
K^+\pi^+\pi^-\pi^0J/\psi$ decays~\cite{x3872_ref5}.  Both decay modes
favor the $C$-parity of $X(3872)$ to be $+1$.

\begin{figure}[htb]
\centering\includegraphics[width=0.82\textwidth]{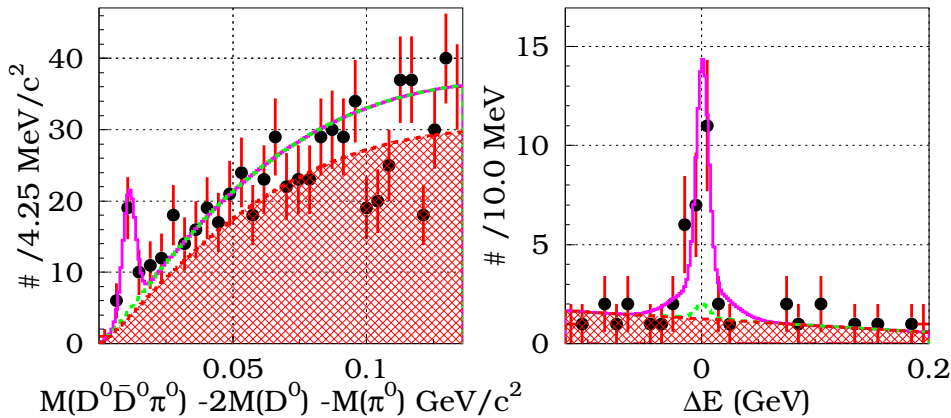}
\caption{Projection of Q-value ($=M_{D^0\bar
D^0\pi^0}-2M_{D^0}-M_{\pi^0}$) ($\Delta E$) when $\Delta E$ (Q-value)
is in the signal region corresponding to $|\Delta E|<25$~MeV
($6$~MeV/$c^2<$ Q-value $<14$~MeV/$c^2$).  The dots with error bars
are data, the hatched histogram corresponds to combinatorial
background; the dashed line indicates the total background and the
solid line is the combined fitting function.}
\label{x3872_signal}
\end{figure}

Recently, the $X(3872)$ state has also been observed in the decay
$X(3872)\to D^0\bar D^0\pi^0$ from $B\to D^0\bar D^0\pi^0K$
decays~\cite{x3872_ref6} (Fig.~\ref{x3872_signal}).  A $414$~fb$^{-1}$
data sample was used for this analysis.  This observation, together
with an angular analysis of $B^+\to K^+\pi^+\pi^-J/\psi$ decays, gives
evidence for $J^{PC}(X(3872))=1^{++}$.

\subsection{\boldmath Observation of $Y(3940)$}

\begin{figure}[htb]
\centering\includegraphics[width=0.4\textwidth]{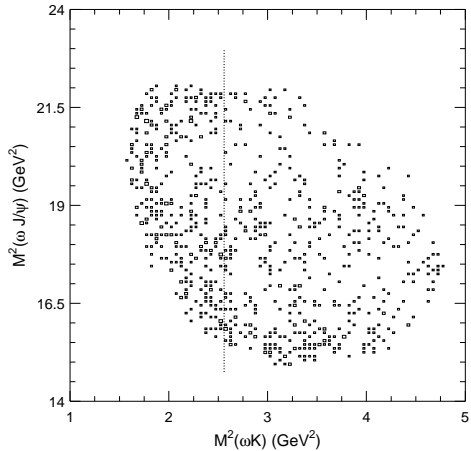}
\caption{Dalitz-plot distribution for $B\to K\omega J/\psi$ candidate
events.  The dotted line indicates the boundary of the
$M(K\omega)>1.6$~GeV/$c^2$ selection requirement.}
\label{y3940_dalitz}
\end{figure}

Another resonant state, denoted as $Y(3940)$, was observed in the
$\omega J/\psi$ system produced in exclusive $B\to K\omega J/\psi$
decays~\cite{y3940_ref1}.  The study was based on a $253$~fb$^{-1}$
data sample that contains $275$ million $B\bar B$ pairs.  To suppress
events of the type $B\to K_XJ/\psi, K_X\to K\omega$, the analysis was
restricted to events in the region
$M(K\omega)>1.6$~GeV/$c^2$~(Fig.~\ref{y3940_dalitz}).

\begin{figure}[htb]
\centering\includegraphics[width=0.72\textwidth]{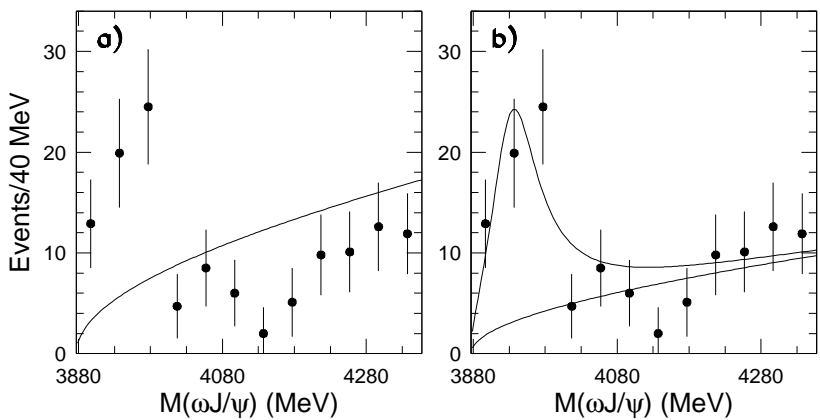}
\caption{$B\to K\omega J/\psi$ signal yields $vs$ $M(\omega J/\psi)$.
The curve in a) indicates the result of a fit that includes only a
phase-space-like threshold function.  The curve in b) shows the result
of a fit that includes an $S$-wave Breit-Wigner resonance term.}
\label{y3940_signal}
\end{figure}

Figure~\ref{y3940_signal} shows the result of fits in bins of the
$\omega J/\psi$ invariant mass.  An enhancement is evident around
$M(\omega J/\psi)=3940$~MeV/$c^2$.  The fit yields a Breit-Wigner
signal yield of $58\pm11$ events with mass $M=3943\pm11$~MeV/$c^2$ and
width $\Gamma=87\pm22$~MeV/$c^2$ (statistical error only).  The
statistical significance of the signal is $8.1~\sigma$.

The product branching fraction was found to be:
\[{\cal B}(B\to KY(3940)){\cal B}(Y(3940)\to\omega
J/\psi)=(7.1\pm1.3\pm3.1)\times10^{-5}.\]

Charmonium states above open charm threshold should dominantly decay
to $D^{(*)}\bar D$ final states.  But these decay modes were not
observed for the $Y(3940)$ resonance.  The properties of the observed
enhancement are similar to those of some of the $c\bar c-gluon$ hybrid
charmonium states that were first predicted in 1978~\cite{xyz_ref2}
and are expected to be produced in $B$ meson decays.  It has been
shown that a general property of these hybrid states is that their
decays to $D^{(*)}\bar D^{(*)}$ meson pairs are forbidden or
suppressed, and the relevant ``open charm'' threshold is
$m_D+m_{D^{**}}\approx4285$~MeV/$c^2$~\cite{y3940_ref2}, where
$D^{**}$ refers to the $J^P=(0,1,2)^+$ charmed mesons.  However, the
predicted masses are substantially higher than the measured value for
$Y(3940)$.

\subsection{\boldmath Observation of $X(3940)$}

\begin{figure}[htb]
\centering\includegraphics[width=0.7\textwidth]{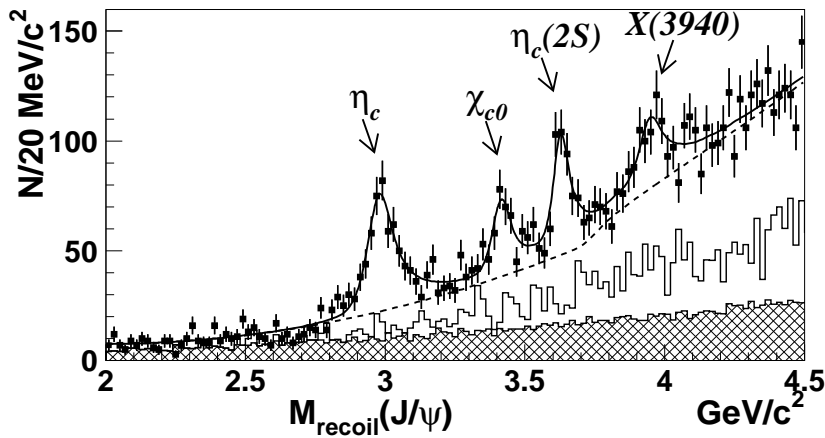}
\caption{The distribution of $M_{\rm recoil}(J/\psi)$ in inclusive
$e^+e^-\to J/\psi X$ events.}
\label{x3940_signal}
\end{figure}

A new charmonium-like state above the $D\bar D$ threshold, which was
denoted as $X(3940)$, has been observed in the process $e^+e^-\to
J/\psi X(3940)$~\cite{x3940_ref1}.  The analysis was based on a
$375$~fb$^{-1}$ data sample.  The signal was searched in the mass of
the system recoiling against the reconstructed particles defined as
\[M_{\rm recoil}(X)=\sqrt{(E_{\rm CM}-E_X^*)^2-p_X^{*\,2}},\]
where $E_X^*$ and $p_X^*$ are the center-of-mass (CM) energy and
momentum.  The $M_{\rm recoil}(J/\psi)$ is shown on
Fig.~\ref{x3940_signal}.  Here, a clean enhancement around
$3.94$~GeV/$c^2$ is seen.  The significance of the $X(3940)$ signal is
$5.0~\sigma$.  The fitted width of the $X(3940)$ state,
$\Gamma=39\pm26$~MeV/$c^2$, is consistent with zero within its large
statistical error.

\begin{figure}[htb]
\centering\includegraphics[width=0.62\textwidth]{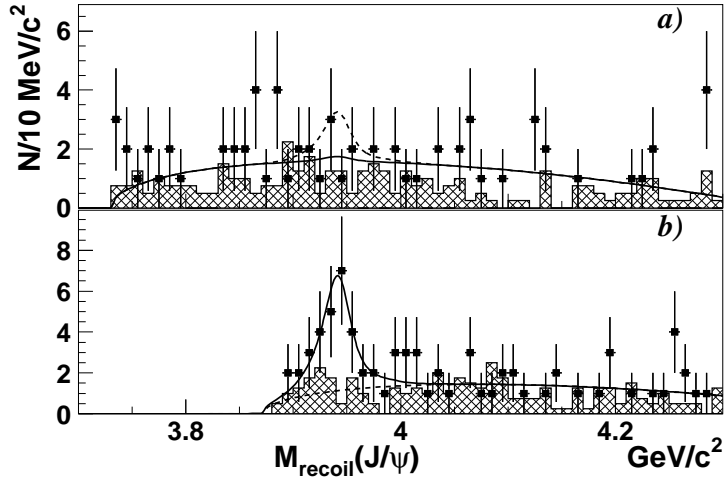}
\caption{The $M_{\rm recoil}(J/\psi)$ distribution for events tagged
and constrained as a) $e^+e^-\to J/\psi D\bar D$, and b) $e^+e^-\to
J/\psi D^*\bar D$.  The hatched histograms correspond to scaled $D$
sidebands.  The solid lines are the results of the fits.  The dashed
lines show: a) the $90\%$ C.L. upper limit on the signal; b) the
background contribution.}
\label{x3940_ddst}
\end{figure}

The new state has a mass above both the $D\bar D$ and $D^*\bar D$
thresholds.  The search for $X(3940)$ decays into $D\bar D$ and
$D^*\bar D$ was performed.  Due to low charm meson reconstruction
efficiency it is not feasible to reconstruct fully $X(3940)\to
D^{(*)}\bar D$.  To increase the efficiency only the prompt $J/\psi$
meson and one $D$ from the $X(3940)$ were reconstructed.  $M_{\rm
recoil}(J/\psi D)$ was constrained to the mass of either $D$ or $D^*$
and the $X(3940)$ signal was searched again in the spectrum of the
mass recoiling against the $J/\psi$ system.  A clear peak has been
seen in the case of $M_{\rm recoil}(J/\psi D)=M(D^*)$ with
significance $5.0~\sigma$, while no signal has been observed in the
case of $M_{\rm recoil}(J/\psi D)=M(D)$ leading to the limit ${\cal
B}(X(3940)\to D\bar D)<41\%$ at $90\%$ C.L. (Fig.~\ref{x3940_ddst}).

The same study was done for the possible decay $X(3940)\to
J/\psi\omega$ and no significant signal was observed, resulting in
${\cal B}(X(3940)\to J/\psi\omega)<26\%$ at $90\%$ C.L.  This means
that $Y(3940)$ and $X(3940)$ are different particles, since they have
different decay modes and different decay widths.  A possible
interpretation of the $X(3940)$ could be $\eta_c(3S)$.

\subsection{\boldmath Observations of $Z(3930)$}

\begin{figure}[htb]
\begin{tabular}{@{}c@{}c}
\put(  10,  0){\includegraphics[width=0.41\textwidth]{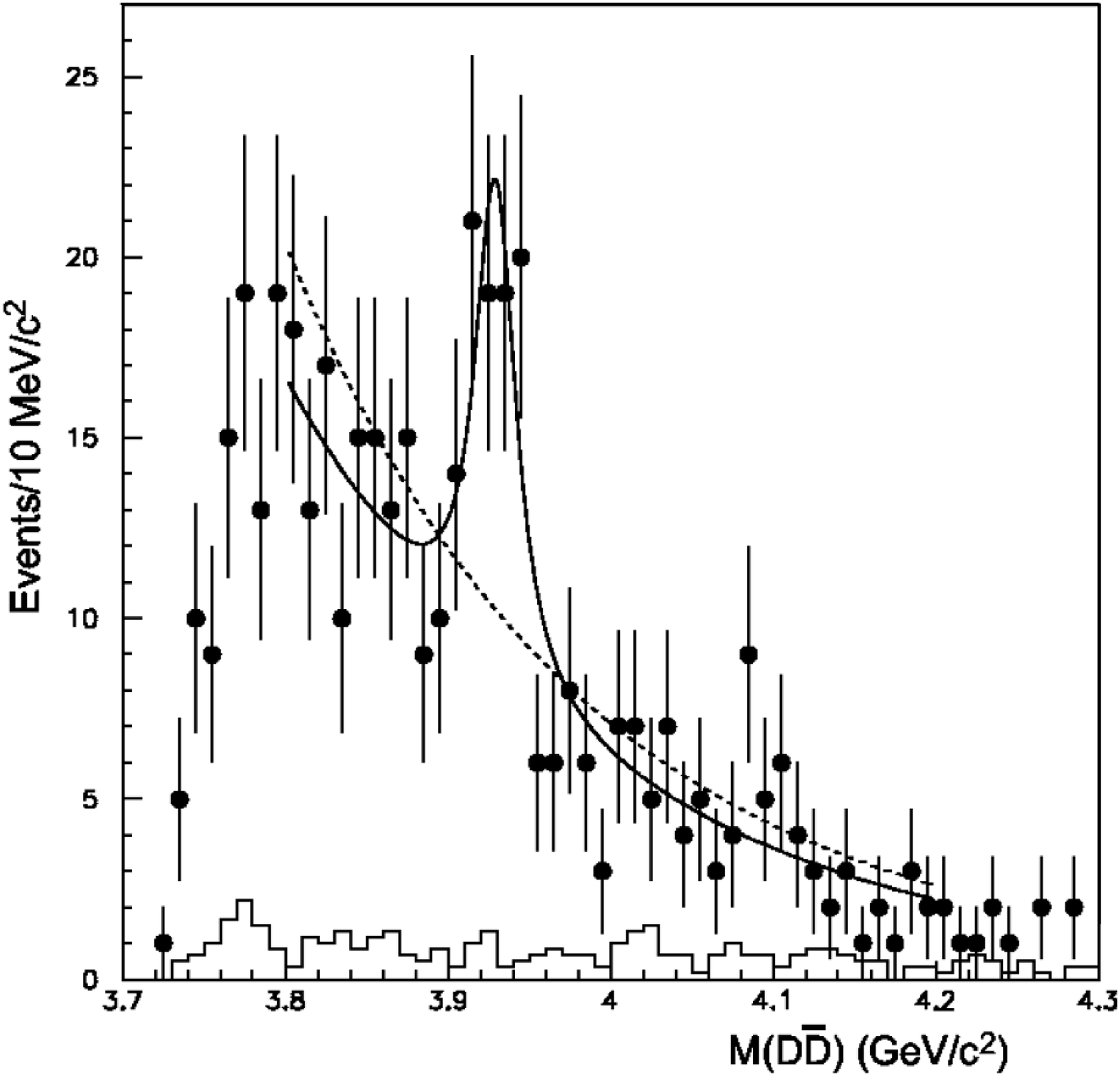}}
\put(  40,160){\bf a)}
\put( 240, -8){\includegraphics[width=0.414\textwidth]{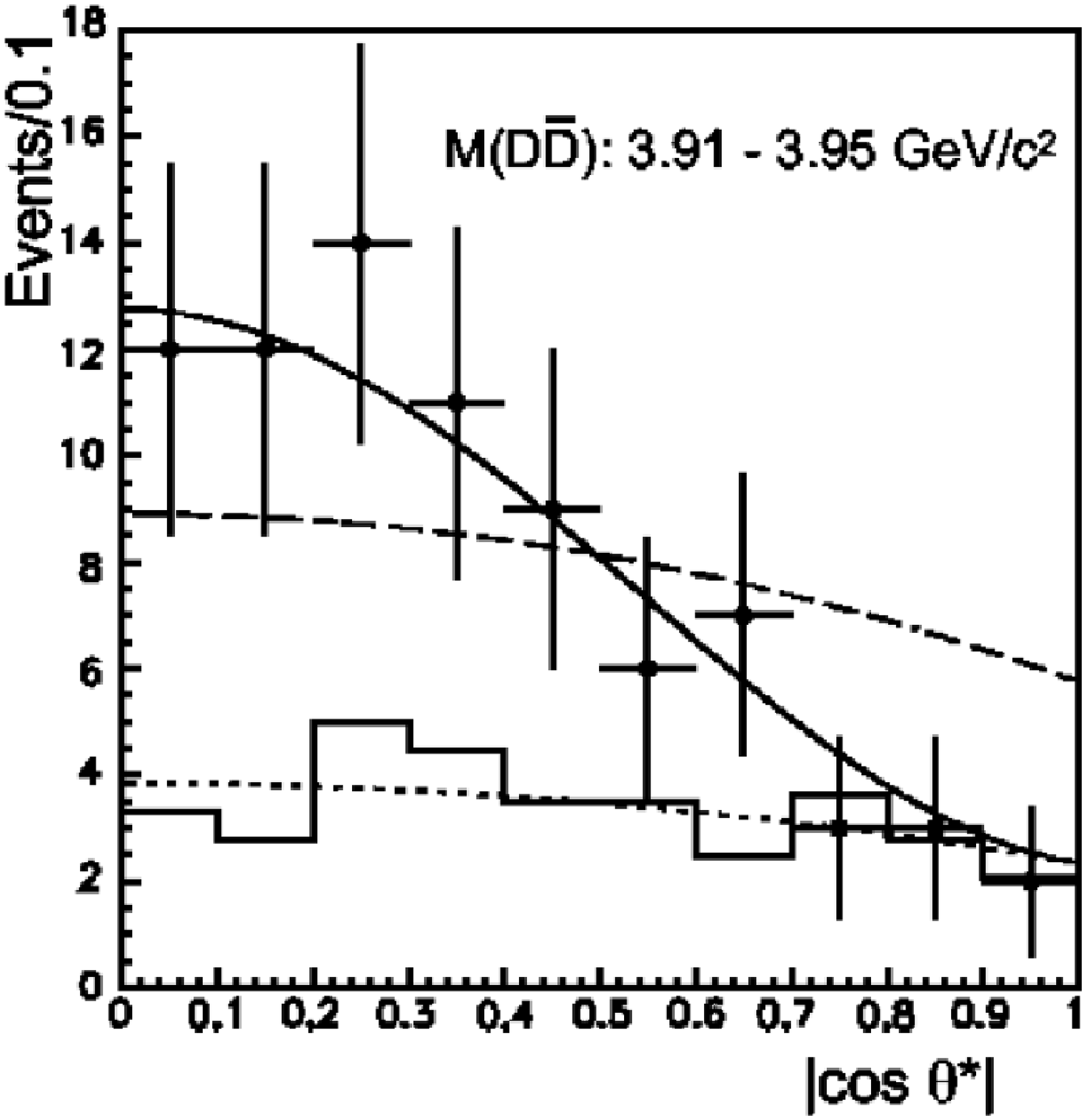}}
\put( 270,160){\bf b)}
\end{tabular}
\caption{a) The $D\bar D$ invariant mass distribution.  The curves
show the fit with (solid) and without (dashed) a resonance component.
The histogram shows the distribution of the events from the
$D$-sidebands.  b) The $|\cos\theta^*|$ distributions in the
($3.91<M(D\bar D)<3.95$)~GeV/$c^2$ region (points with error bars) and
background scaled from the $M(D\bar D)$ sidebands (solid histogram).
The solid and dashed curves are expected distributions for the spin
two (helicity two) and spin zero hypotheses, respectively, and contain
the non-peaking background also shown separately by the dotted curve.}
\label{z3930_signal}
\end{figure}

A search for the $\chi_{cJ}^\prime$ ($J=0$ or $2$) states and other
$C$-even charmonium states in the mass region of $3.73-4.3$~GeV/$c^2$
produced via the process $\gamma\gamma\to D\bar D$ was performed using
$395$~fb$^{-1}$ data sample~\cite{z3930_ref1}.  The two-photon process
$e^+e^-\to e^+e^-D\bar D$ was studied in the ``zero-tag'' mode, where
neither the final state electron nor positron is detected, and the
$D\bar D$ system has very small transverse momentum: $P_t(D\bar
D)<0.05$~GeV/$c$.  In Fig.~\ref{z3930_signal}a the $M(D\bar D)$
invariant mass distribution is shown for the combined $D^0\bar D^0$
and $D^+D^-$ channels.

The fit results for the peak near $3.93$~GeV/$c^2$ for the resonance
mass, width and total yield of the resonance are
$M=3929\pm5$~MeV/$c^2$, $\Gamma=29\pm10$~MeV and $64\pm18$ events,
respectively.  The statistical significance of the peak is
$5.3~\sigma$.

Fig.~\ref{z3930_signal}b shows the event yields in the
$3.91-3.95$~GeV/$c^2$ region versus $|\cos\theta^*|$.  The curves show
the expectations for cases of $J=2$ (solid line) and $J=0$ (dashed
line).  The data significantly favor a spin two assignment over spin
zero.  Using the number of observed events, the product of the
two-photon decay width and $D\bar D$ branching fraction was determined
to be
\[\Gamma_{\gamma\gamma}(Z(3930)){\cal B}(Z(3930)\to D\bar
D)=0.18\pm0.05\pm0.03{\rm~keV}.\]
The measured properties are consistent with expectations for the
previously unseen $\chi_{c2}^\prime$ charmonium state.

\section{\boldmath First results from $\Upsilon(5S)$ data}

The possibility of studying $B_s$ decays at very high luminosity
$e^+e^-$ colliders running at the energy of the $\Upsilon(5S)$
resonance has been discussed in several theoretical
papers~\cite{5s_ref1,5s_ref2,5s_ref3}.  To test the experimental
feasibility of $B_s$ studies in $\Upsilon(5S)$ events, a sample of
$1.86$~fb$^{-1}$ was collected by the Belle detector over three days
in June 2005.  Another data sample of $3.67$~fb$^{-1}$ taken at a CM
energy $60$~MeV below the $\Upsilon(4S)$ was used in this analysis to
evaluate continuum contributions.

In the energy region of the $\Upsilon(5S)$, the hadronic events can be
classified into three physics categories: $u\bar u, d\bar d, s\bar s,
c\bar c$ continuum events, $b\bar b$ continuum events and
$\Upsilon(5S)$ events.  All $b\bar b$ events (including those from
$\Upsilon(5S)$) are expected to hadronize in one of the following
final states: $B\bar B, B\bar B^*, B^*\bar B, B^*\bar B^*, B\bar B\pi,
B\bar B^*\pi, B^*\bar B\pi, B^*\bar B^*\pi, B\bar B\pi\pi, B_s^0\bar
B_s^0, B_s^0\bar B_s^*, B_s^*\bar B^0$ or $B_s^*\bar B_s^*$.  Here $B$
denotes a $B^0$ or a $B^+$ meson and $\bar B$ denotes a $\bar B^0$ or
a $B^-$ meson.  The excited states decay to their ground states via
$B^*\to B\gamma$ and $B_s^*\to B_s^0\gamma$~\cite{PDG}.

Using the numbers of hadronic events in the $\Upsilon(5S)$ and
continuum data samples, the number of $b\bar b$ events was measured to
be $N_{b\bar b}(\Upsilon(5S))=(5.61\pm0.03(\rm stat)\pm0.29(\rm
syst))\times10^5$.  It corresponds to $N_{b\bar b}(\Upsilon(5S))/{\rm
fb}^{-1}=(3.02\pm0.15)\times10^5$~\cite{5s_ref4}, which is consistent
with the CLEO measurement $N_{b\bar b}(\Upsilon(5S))/{\rm
fb}^{-1}=(3.10\pm0.52)\times10^5$.

\subsection{Inclusive $\Upsilon(5S)\to D_sX$ study}

\begin{figure}[htb]
\vspace{.3cm}
\centering
\begin{tabular}{@{}l@{}r}
\includegraphics[width=0.41\textwidth]{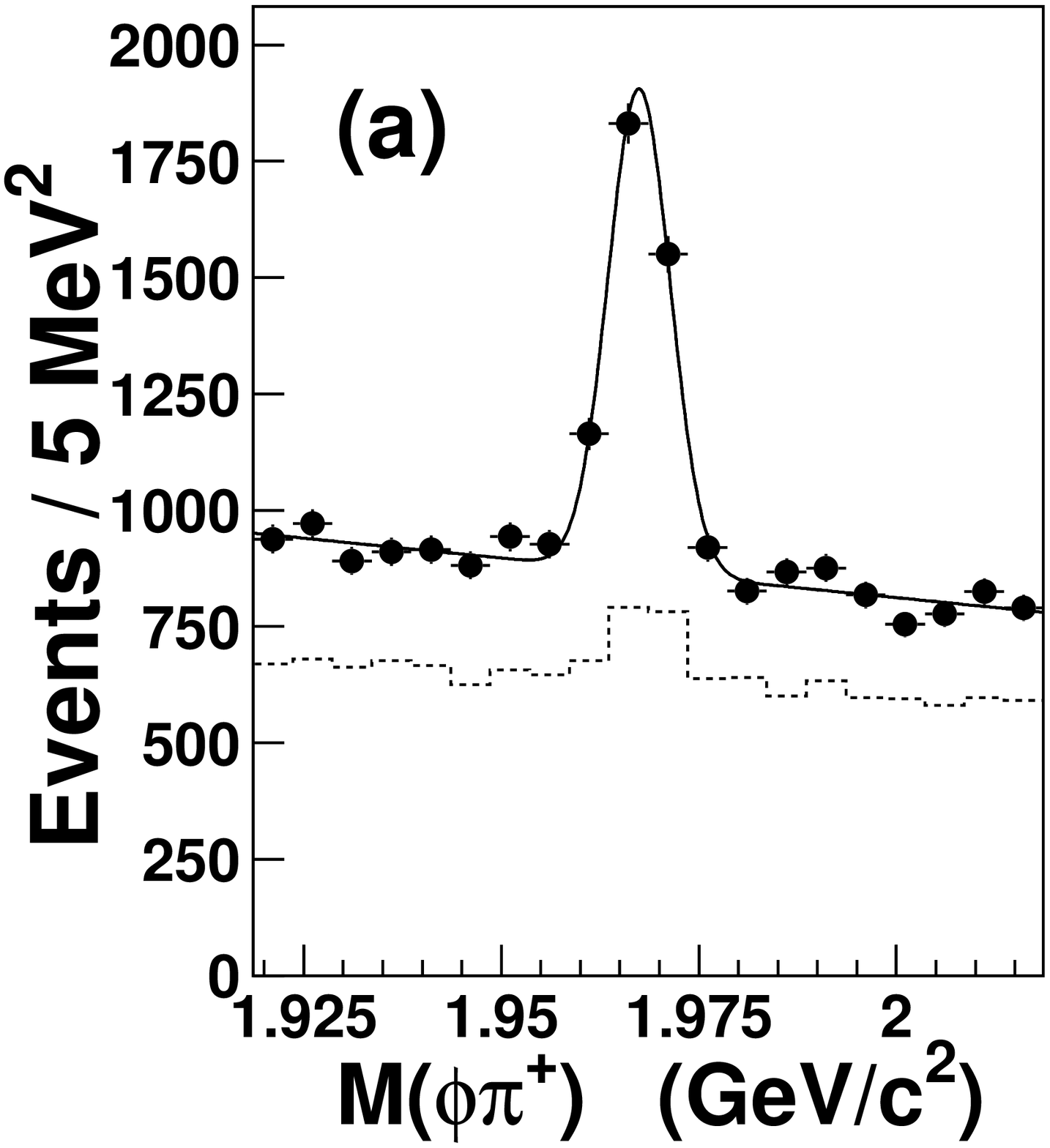}
\includegraphics[width=0.41\textwidth]{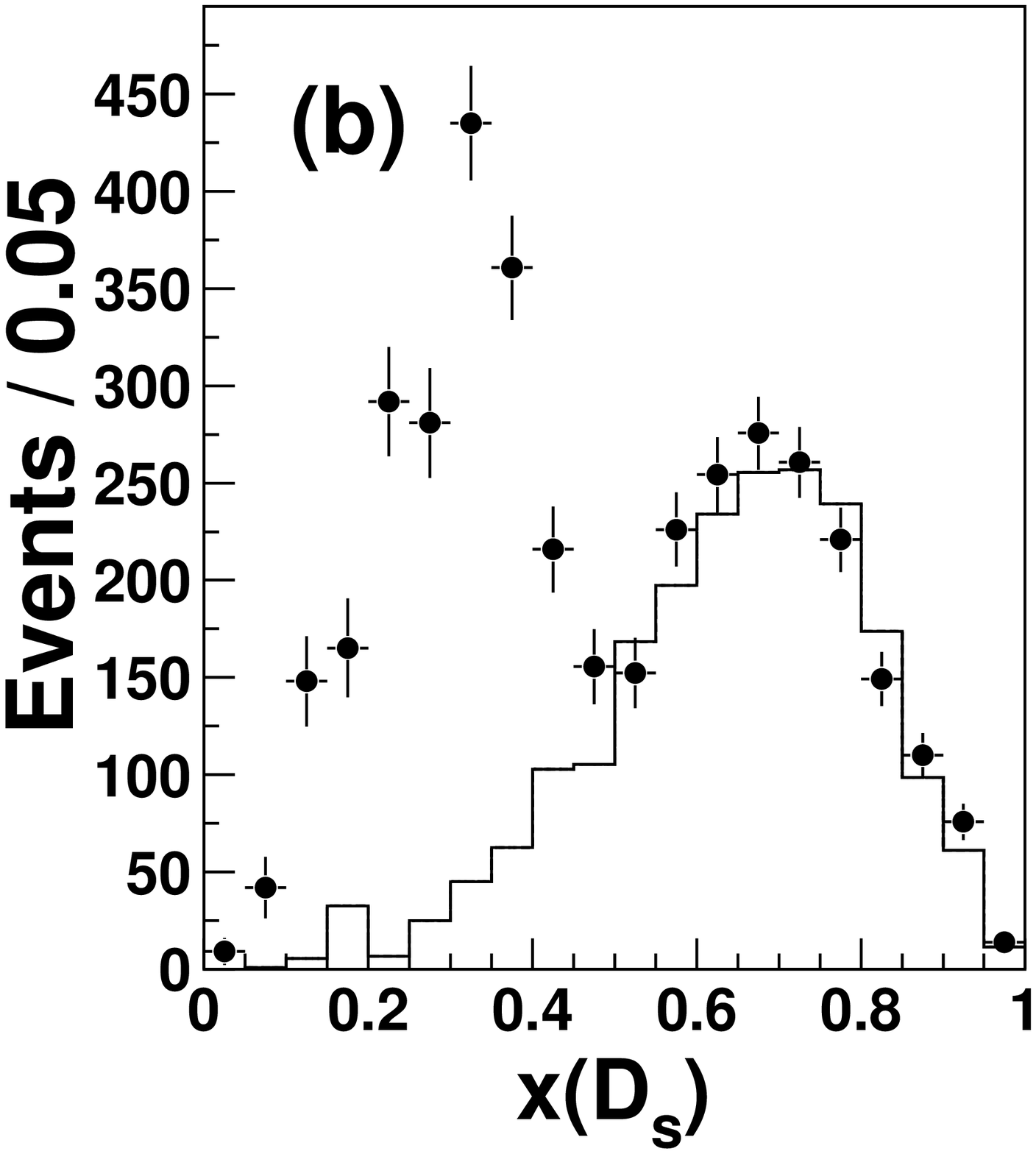}
\end{tabular}
\caption{The $D_s$ signal in the region $x(D_s)<0.5$ (a) and the $D_s$
normalized momentum $x(D_s)$ (b).  The points with error bars are the
$\Upsilon(5S)$ data, while the histograms show the normalized
continuum.}
\label{5s_xds}
\end{figure}

To measure the fraction $f_s$ of $B_s^{(*)}\bar B_s^{(*)}$ events over
the total number of $b\bar b$ events, the inclusive $D_s$ production
was studied.  Finally, $b\bar b$ events from $\Upsilon(5S)$ data can
decay either to $B_s\bar B_s+X$ or to $B\bar B+X$.  The total
inclusive branching fraction of $D_s$ production can therefore be
expressed as:
\[{\cal B}(\Upsilon(5S)\to D_sX)/2=f_s\times{\cal B}(B_s\to
D_sX)+(1-f_s)\times{\cal B}(B\to D_sX),\] where ${\cal B}(B_s\to
D_sX)$ is theoretically predicted to be $(92\pm11)\%$~\cite{5s_ref5},
and ${\cal B}(B\to D_sX)=(8.7\pm1.2)\%$~\cite{PDG,5s_ref5} is well
measured at the $\Upsilon(4S)$.  The $D_s$ was reconstructed in the
cleanest mode $\phi\pi, \phi\to K^+K^-$.  The $D_s$ signal in the
$\Upsilon(5S)$ and continuum data samples with $x(D_s)=p(D_s)/p_{\rm
max}(D_s)<0.5$ are shown in Fig.~\ref{5s_xds}a.  The normalized
momentum distributions $x(D_s)$ are shown in Fig.~\ref{5s_xds}b for
the same data samples.
The excess of events in the region $x(D_s)<0.5$ corresponds to
inclusive $D_s$ production in $b\bar b$ events.  Finally, the
inclusive branching fraction ${\cal B}(\Upsilon(5S)\to
D_sX)/2=(23.6\pm1.2\pm3.6)\%$ was obtained.  Using this result the
value of $f_s$ was measured to be $f_s=(17.9\pm1.4\pm4.1)\%$,
consistent with CLEO result $f_s=(16.0\pm2.6\pm5.8)\%$.  In the same
way, but using $D^0$ instead of $D_s$, it was measured ${\cal
B}(\Upsilon(5S)\to D^0X)/2=(53.8\pm2.0\pm3.4)\%$ and
$f_s=(18.1\pm3.6\pm7.5)\%$~\cite{5s_ref4}.

\subsection{Fully reconstructed exclusive $B_s$}

\begin{figure}[htb]
\includegraphics[width=1.03\textwidth]{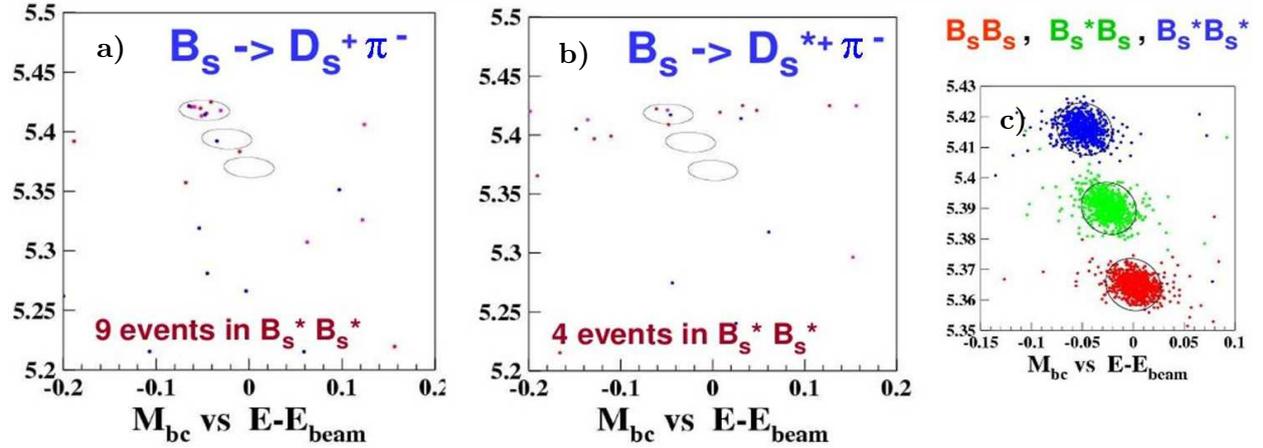}
\put(-433,147){\bf a)}
\put(-260,145){\bf b)}
\put(-95,120){\bf c)}
\caption{$M_{\rm bc}$ $vs$ $\Delta E$ scatter plot for a) $B_s\to
D_s^+\pi^-$, b) $B_s\to D_s^{*+}\pi^-$ for the data, and c) signal MC
events.}
\label{5s_signal}
\end{figure}

As mentioned above, the $\Upsilon(5S)$ meson can decay to the
following $B_s$ decay modes: $B_sB_s, B_s^*B_s, B_s^*B_s^*$, where
$B_s^*\to B_s\gamma$.  MC simulation shows that if the photon(s)
is(are) lost, both $\Delta E$ and $M_{\rm bc}$ signal distributions
are shifted, and the three types of $\Upsilon(5S)$ decays can be well
distinguished by reconstructing only $B_s$ mesons without $\gamma$
reconstruction (Fig.~\ref{5s_signal}c).  The $B_s$ was reconstructed
in the cleanest decay modes $B_s\to D_s^-\pi^+$ and $B_s\to
D_s^{*-}\pi^+$ with $D_s^{*-}\to D_s^-\gamma$ and $D_s^-\to\phi\pi^-$.
A clear signal was observed in the $B_s^*B_s^*$ channel, while no
signal was observed in $B_s^*B_s$ and $B_sB_s$ (Fig.~\ref{5s_signal}a
and b).
  Taking into account the
number of $B_s$ mesons obtained from inclusive analysis, the branching
fraction for the decay mode $B_s\to D_s^-\pi^+$ was measured to be:
\[{\cal B}(B_s\to D_s^-\pi^+)=(0.68\pm0.22\pm0.16)\%.\]

The fraction of $B_s^*B_s^*$ events over all $B_s^{(*)}B_s^{(*)}$
events was measured to be\\
$N(B_s^*B_s^*)/N(B_s^{(*)}B_s^{(*)})=(94^{+6}_{-9})$~\cite{5s_ref6}.
The Potential models predict the dominance of $B_s^*B_s^*$ over
$B_s^*B_s$ and $B_sB_s$ channels, but not as strong as what was
observed.

Combining all studied decay modes, $B_s^*$ and $B_s$ masses were
measured to be: $M(B_s^*)=5418\pm1\pm3$~MeV/$c^2$ and
$M(B_s)=5370\pm1\pm3$~MeV/$c^2$~\cite{5s_ref6}.  The obtained $B_s$
mass is in agreement with the recent CDF measurement
$M(B_s)=5366.0\pm0.8$~MeV/$c^2$.

\section{Summary}

The first evidence for the decay $B\to\tau\nu$ has been reported by
Belle.  This is the first direct measurement of $f_B$.

Several new resonances have been observed by Belle: $X(3872)$,
$Y(3940)$, $X(3940)$, $Z(3930)$.  The first two can not be ascribed to
the expected particle states.  The possible explanations include
$D^*\bar D$ molecules, $c\bar c$-gluon hybrids or tetraquarks.

Results from the $\Upsilon(5S)$ engineering run are very promising.
Even with only a small amount of data some significant (preliminary)
results were obtained.


\begin{thebibliography}{99}

\bibitem{beldetec} Belle Collaboration, A.~Abashian {\it et al.},
Nucl. Instr. and Meth. A{\bf 479}, 117-232 (2002).

\bibitem{KEKB} S.~Kurokawa and E.~Kikutani, Nucl. Instr. and Meth. A
{\bf 499}, 1 (2003).

\bibitem{cc} The inclusion of charge conjugate modes is implied
throughout this Letter.

\bibitem{PDG} S.~Eidelman {\it et al.} (Particle Data Group),
Phys. Lett. B {\bf 592}, 1 (2004).

\bibitem{tau_ref1} E.~Barberio {\it et al.} (Heavy Flavor Averaging Group),
hep-ex/0603003.

\bibitem{tau_ref2} A.~Gray {\it et al.} (HPQCD Collaboration),
Phys. Rev. Lett. {\bf 95}, 212001 (2005).

\bibitem{tau_ref3} W.S.~Hou, Phys. Rev. D {\bf 48}, 2342 (1993).

\bibitem{tau_ref4} B.~Aubert {\it et al.} (BABAR Collaboration),
Phys. Rev. D {\bf 73}, 057101 (2006).

\bibitem{tau_ref5} K.~Ikado {\it et al.} (Belle Collaboration),
hep-ex/0604018.

\bibitem{xyz_ref1} M.B.~Voloshin and L.B.~Okun, JETP Lett. {\bf 23},
333 (1976); M.~Bander, G.L.~Shaw and P.~Thomas, Phys. Rev. Lett. {\bf
36}, 695 (1977); A.~De~Rujula, H.~Georgi and S.L.~Glashow,
Phys. Rev. Lett. {\bf 38}, 317 (1977); N.A.~Tornqvist, Z. Phys. C {\bf
61}, 525 (1994); and A.V.~Manohar and M.B.~Wise, Nucl.Phys. B {\bf
339}, 17 (1993).

\bibitem{xyz_ref2} D.~Horn and J.~Mandula, Phys. Rev. D {\bf 17}, 898
(1978).

\bibitem{x3872_ref1} S.K.~Choi {\it et al.} (Belle Collaboration),
Phys. Rev. Lett. {\bf 91}, 262001 (2003).

\bibitem{x3872_ref2} D.~Acosta {\it et al.} (CDF-II Collaboration),
Phys. Rev. Lett. {\bf 93}, 072001 (2004); V.M.~Abazov {\it et al.} (D0
Collaboration), Phys. Rev. Lett. {\bf 93}, 162002 (2004); and B.Aubert
{\it et al.} (BABAR Collaboration), Phys. Rev. D {\bf 71}, 071103
(2005).

\bibitem{x3872_ref3} S.L.~Olsen, Int. J. Mod. Phys. A{\bf 20}, 240-249
(2005) and K.~Abe {\it et al.} (Belle Collaboration), hep-ex/0408116.

\bibitem{x3872_ref4} N.A.~Tornqvist, Phys. Lett.  B {\bf 590}, 209
  (2004); E.S.~Swanson, Phys.  Lett.  B {\bf 588}, 189 (2004);
  F.E.~Close and P.R.~Page, Phys.  Lett. B {\bf 578}, 119 (2003);
  S.~Pakvasa and M.~Suzuki, Phys. Lett. B {\bf 579}, 67 (2004);
  C.-Y.~Wong, Phys. Rev. C {\bf 69}, 055202 (2004); and E.~Braaten and
  M.~Kusunoki, Phys. Rev. D {\bf 69}, 114012 (2004).

\bibitem{x3872_ref5} K.~Abe, {\it et al.} (Belle Collaboration)
hep-ex/0505037.

\bibitem{x3872_ref6} G.~Gokhroo {\it et al.} (Belle Collaboration),
hep-ex/0606055.

\bibitem{y3940_ref1} S.K.~Choi {\it et al.} (Belle Collaboration),
Phys. Rev. Lett. {\bf 94}, 182002 (2005).

\bibitem{y3940_ref2} N.~Isgur, R.~Kokoski and J.~Paton,
Phys. Rev. Lett. {\bf 54}, 869 (1985).

\bibitem{x3940_ref1} K.~Abe {\it et al.} (Belle Collaboration),
hep-ex/0507019.

\bibitem{z3930_ref1} K.~Abe {\it et al.} (Belle Collaboration),
Phys. Rev. Lett. {\bf 96}, 082003 (2006).

\bibitem{5s_ref1} A.F.~Falk and A.A.~Petrov, Phys. Rev. Lett. {\bf
85}, 252 (2000).

\bibitem{5s_ref2} S.~Petrak, SLAC-PREPRINT-2001-041.

\bibitem{5s_ref3} D.~Atwood and A.~Soni, Phys. Lett. B {\bf 533}, 37
(2002).

\bibitem{5s_ref4} A.~Drutskoy {\it et al.} (Belle Collaboration),
hep-ex/0608015.

\bibitem{5s_ref5} M.~Artuso {\it et al.} (CLEO Collaboration),
Phys. Rev. Lett. {\bf 95}, 261801 (2005).

\bibitem{5s_ref6} K.~Abe {\it et al.} (Belle Collaboration),
hep-ex/0610003.

\end{thebibliography}
\end{document}